\documentclass[12pt]{article}
 \usepackage{epsfig}
 \def\be{\begin{equation}}
 \def\ee{\end{equation}}
 \def\bea{\begin{eqnarray}}
 \def\eea{\end{eqnarray}}
 \usepackage{graphicx}

 \catcode`\@=11
 \def\lsim{\mathrel{\mathpalette\@versim<}}
 \def\gsim{\mathrel{\mathpalette\@versim>}}
 \def\@versim#1#2{\vcenter{\offinterlineskip
 \ialign{$\m@th#1\hfil##\hfil$\crcr#2\crcr\sim\crcr } }}
 \catcode`\@=12

 \catcode`@=12
\begin{document}
 \thispagestyle{empty}
 \begin{flushright}
 UCRHEP-T570\\
 August 2016\
 \end{flushright}
 \vspace{0.2in}
 \begin{center}
 {\LARGE \bf Connecting Radiative Neutrino Mass,\\ Neutron-Antineutron
Oscillation,\\ Proton Decay, and Leptogenesis\\ through Dark Matter\\}
 \vspace{0.8in}
 {\bf Pei-Hong Gu$^1$, Ernest Ma$^2$, and Utpal Sarkar$^3$\\}
 \vspace{0.2in}
{\sl $^1$ Department of Physics and Astronomy, Shanghai Jiao Tong University,\\
800 Dongchuan Road, Shanghai 200240, China\\}
\vspace{0.1in}
 {\sl $^2$ Physics \& Astronomy Department and Graduate Division,\\
 University of California, Riverside, California 92521, USA\\}
\vspace{0.1in}
{\sl $^3$ Physics Department, Indian Institute of Technology,\\
Kharagpur 721302, India\\}
 \end{center}
 \vspace{0.8in}

\begin{abstract}\
The scotogenic mechanism for radiative neutrino mass is generalized
to include neutron-antineutron oscillation as well as proton decay.
Dark matter is stabilized by extending the notion of lepton parity
to matter parity. Leptogenesis is also a possible byproduct. This
framework unifies the description of all these important, but seemingly
unrelated, topics in physics beyond the standard model of particle
interactions.
\end{abstract}

 \newpage
 \baselineskip 24pt

\noindent \underline{\it Introduction}~:~\\
The standard model (SM) of quarks and leptons is based on the well-tested
$SU(3)_C \times SU(2)_L \times U(1)_Y$ gauge symmetry. It admits two
well-known accidental symmetries, baryon number $B$ and lepton number $L$,
which are known to be conserved as far as present experimental limits are
concerned. Of course, if neutrino masses are confirmed as Majorana
from neutrinoless double beta decay in the near future, then $L$
should be downgraded to just $(-1)^L$, i.e. lepton parity $P_L$. This is
actually an important concept, because dark matter may be stabilized
by the proper extension of $P_L$ to physics beyond the SM~\cite{m15}.
It would also tell us that $P_L$ may be the true symmetry of a
complete theory, whereas the conservation of $L$ only holds in the absence
of neutrino masses.

Let us now consider $B$. Is there a possible clue that it is not the
true symmetry of a complete theory? The analog to Majorana neutrino
mass is then neutron-antineutron ($n-\bar{n}$) oscillation. If proven to exist,
$B$ would be downgraded to $(-1)^{3B}$, i.e. baryon parity $P_B$. What about
proton decay? If it exists, then the final product must contain a
lepton, e.g. $p \to \pi^0 e^+$ or $p \to \pi^+ \nu (\bar{\nu})$.
This would violate both lepton parity and baryon parity.  It
may however be accommodated by combining lepton parity with baryon parity
to form matter parity, i.e. $P_M = (-1)^{3B+L}$.

In this paper, we assume that the true symmetry of
a complete theory beyond the SM is $P_M$.  However, $P_L$ and $P_B$ are
respected by all dimension-four and dimension-three terms of the Lagrangian,
broken only to $P_M$ by a unique dimension-two term.  With the present available
experimental accuracy, the separate conservation of $B$ and $L$ holds.
To confirm our hypothesis, it would take future extraordinary discoveries,
i.e. neutrinoless double beta decay (for $P_L$), neutron-antineutron
oscillation (for $P_B$), and proton decay (for $P_M$). Nevertheless, there is
already a theoretical framework for connecting all of these
phenomena. It is the scotogenic mechanism (from the Greek {\it scotos}
meaning darkness), invented 10 years ago~\cite{m06}.

\noindent \underline{\it Scotogenic neutrino mass}~:~\\
The scotogenic mechanism  was applied to obtaining one-loop Majorana neutrino
masses as shown in Fig.~1.
\begin{figure}[htb]
\vspace*{-3cm}
\hspace*{-3cm}
\includegraphics[scale=1.0]{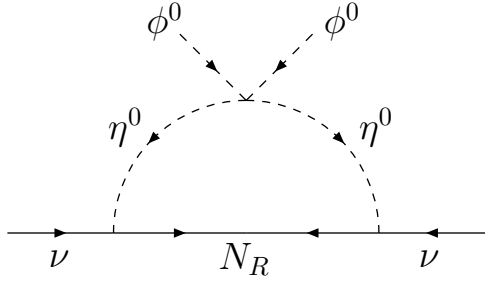}
\vspace*{-21.5cm}
\caption{One-loop $Z_2$ scotogenic neutrino mass.}
\end{figure}
The scalar doublet $(\eta^+,\eta^0)$ is odd under $P_L$ to distinguish it
from the SM Higgs doublet $(\phi^+,\phi^0)$ which is even. The three
neutrinos $\nu_L$ are odd under $P_L$, and the three singlet
neutral fermions $N_R$ are even.  The latter have allowed Majorana masses
$m_N$, forming thus three Majorana fermions $N=N_R+N_R^c$.  Note that $N_R$
are not the right-handed neutrinos which would be odd under $P_L$.  This
assignment is equivalent~\cite{m15} to having odd dark parity for $\eta$
and $N_R$, and even dark parity for $\nu_L$ and $\phi$, using the conserved
product $P_L(-1)^{2j}$, where $j$ is the spin angular momentum of the particle.

\noindent \underline{\it Scotogenic neutron-antineutron oscillation}~:~\\
The scotogenic analog for $n-\bar{n}$ oscillation is actually very simple.
Add to the SM two color scalar triplets, one with even $P_B$ and the other
with odd $P_B$ as follows:
\begin{equation}
\delta \sim (3,1,-1/3;+), ~~~ \xi \sim (3,1,-1/3;-).
\end{equation}
The resulting allowed interactions are
$d_R N_R \xi^*$, $(\delta^* \xi)^2$, and $u_{L,R} d_{L,R} \delta$.
Hence $\delta$ is a scalar diquark, and
$n-\bar{n}$ oscillation is generated as shown in Fig.~2.
Note that $N_R$ is again used because it has even $P_B$ as well as $P_L$,
and the scalar $\xi$ inside the loop has odd dark parity.  The new
particles of this model are listed in Table 1, together with two other real
scalar singlets $\chi_{1,2}$ to be discussed later.  The dark parity
$P_D$ is simply defined as $P_M (-1)^{2j}$.
\begin{figure}[htb]
\vspace*{-3cm}
\hspace*{-3cm}
\includegraphics[scale=1.0]{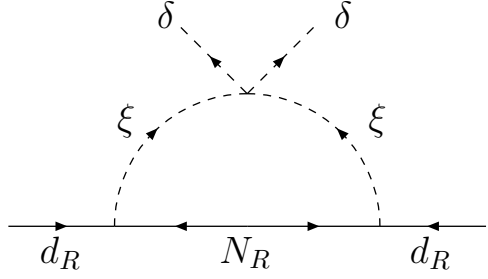}
\vspace*{-21.5cm}
\caption{One-loop $Z_2$ scotogenic $n-\bar{n}$ oscillation.}
\end{figure}

\noindent \underline{\it Scotogenic proton decay}~:~\\
Let us now bisect Figs.~1 and 2 and try to join the two different halves.
The quartic couplings $(\delta^* \xi)(\bar{\phi}^0\eta^0)$ and
$(\delta^* \xi)({\phi}^0\bar{\eta}^0)$ are forbidden by $P_L$ and by
$P_B$.  However they may be induced by the trilinear couplings
$\delta^* \xi \chi_1$, $\bar{\phi}^0\eta^0 \chi_2$, and
${\phi}^0\bar{\eta}^0 \chi_2$, which respect both $P_L$ and $P_B$.
\begin{table}[htb]
\caption{Particle content of proposed model.}
\begin{center}
\begin{tabular}{|c|c|c|c|c|c|c|c|}
\hline
Particle & $SU(3)_C$ & $SU(2)_L$ & $U(1)_Y$ & $P_L$ & $P_B$ & $P_M$ & $P_D$ \\
\hline
$(u,d)_{L}$ & 3 & 2 & 1/6 & $+$ & $-$ & $-$ & $+$ \\
$u_{R}$ & $3$ & 1 & $2/3$ & $+$ & $-$ & $-$ & $+$ \\
$d_{R}$ & $3$ & 1 & $-1/3$ & $+$ & $-$ & $-$ & $+$ \\
$(\nu,l)_{L}$ & 1 & 2 & $-1/2$ & $-$ & $+$ & $-$ & $+$ \\
$l_{R}$ & 1 & 1 & $-1$ & $-$ & $+$ & $-$ & $+$ \\
$N_R$ & 1 & 1 & 0 & $+$ & $+$ & $+$ & $-$ \\
\hline
$(\phi^+,\phi^0)$ & 1 & 2 & 1/2 & + & + & + & + \\
$(\eta^+,\eta^0)$ & 1 & 2 & 1/2 & $-$ & $+$ & $-$ & $-$ \\
$\delta$ & 3 & 1 & $-1/3$ & $+$ & $+$ & $+$ & $+$ \\
$\xi$ & 3 & 1 & $-1/3$ & $+$ & $-$ & $-$ & $-$ \\
\hline
$\chi_1$ & 1 & 1 & 0 & $+$ & $-$ & $-$ & $-$ \\
$\chi_2$ & 1 & 1 & 0 & $-$ & $+$ & $-$ & $-$ \\
\hline
\end{tabular}
\end{center}
\end{table}
The dimension-two mass-squared term $m^2_{12} \chi_1 \chi_2$ is then
inserted to
break $P_L$ and $P_B$ softly to $P_M = P_L P_B$.  The resulting diagrams
are shown in Figs.~3 and 4.
\begin{figure}[htb]
\vspace*{-3cm}
\hspace*{-3cm}
\includegraphics[scale=1.0]{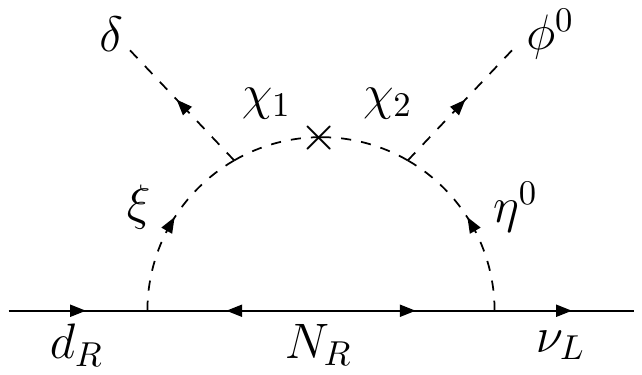}
\vspace*{-21.5cm}
\caption{One-loop $Z_2$ scotogenic $n \to \nu$ transition.}
\end{figure}
\begin{figure}[htb]
\vspace*{-3cm}
\hspace*{-3cm}
\includegraphics[scale=1.0]{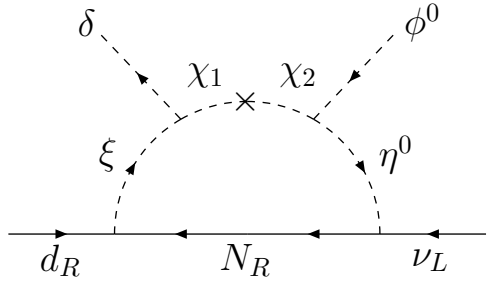}
\vspace*{-21.5cm}
\caption{One-loop $Z_2$ scotogenic $n \to \bar{\nu}$ transition.}
\end{figure}
Both processes conserve $P_M$, whereas $n \to \nu$ conserves
$B+L$ in Fig.~3, and $n \to \bar{\nu}$ conserves $B-L$ in Fig.~4.
It will be shown later that the integral associated with Fig.~4 is
negligible compared to that of Fig.~3.  Hence proton decay proceeds
mainly via $p \to \pi^+ \nu$, thereby conserving $B+L$~\cite{v95,bm12},
instead of the usual $B-L$.  However, since $\nu$ cannot be distinguished
from $\bar{\nu}$ in practice, this prediction cannot be tested.
In this scenario, we assume $\xi$ to be heavier than $N$,
so $\xi$ decays to $N$ + $d$. We also assume that the lightest $N$ is
heavier than $\eta$, so that its decay to $\eta^+ l^-$ and $\eta^- l^+$
may generate a lepton asymmetry~\cite{fy1986}, which gets converted to a
baryon asymmetry through the sphalerons~\cite{krs1985} before the
electroweak phase transition is over.  The dark-matter candidate
is thus either the real or imaginary component of $\eta^0$~\cite{dm78}.
For some recent studies on this possibility, see for example
Refs.~\cite{atyy14,ikr15,dku16}.

\noindent \underline{\it Evoution of $B$ and $L$ symmetries}~:~\\
In our scenario the heaviest particle is $\xi$.  For convenience we also
assume $\chi_{1,2}$ to be at this mass scale.  They may however be much
lighter and not affect our following discussion.  As the Universe cools
below $m_\xi$, the effective theory (minus $\xi$ and possibly $\chi_{1,2}$)
gains the symmetry $B$ in all its dimension-four terms, whereas the
dimension-five
term $(\delta^* d_R)^2$ breaks $B$ to $P_B$.  The $(\delta^* d_R)(\phi^0 \nu
- \phi^+ l^-)$ and $(\delta^* d_R)(\bar{\phi}^0 \bar{\nu} - \phi^- l^+)$
terms break $P_L$ and $P_B$ to $P_M$.  The next heaviest particles
are $N_{1,2,3}$.  As the Universe further cools below their masses,
the effective theory gains also the symmetry $L$ in all its dimension-four
terms, whereas the dimension-five term $(\phi^0 \nu - \phi^+ l^-)^2$ breaks
$L$ to $P_L$.  Meanwhile, the decay $N \to l^\pm \eta^\mp$ has created
a lepton asymmetry and is being converted by sphalerons to the observed baryon
asymmetry of the Universe.  Finally at the electroweak scale, the particle
content of our proposal is that of the SM plus the dark scalar doublet
$(\eta^+,\eta^0)$, and perhaps also the scalar diquark $\delta$.  If $m_\delta$
is much heavier, then the $n-\bar{n}$ oscillation effective operators
$(u_R d_R) d_R d_R (u_R d_R)$, $(u_L d_L) d_R d_R (u_L d_L)$,
$(u_R d_R) d_R d_R (u_L d_L)$ are
dimension-nine, and the proton decay effective operators
$(u_R d_R) d_R (\bar{\phi}^0 \bar{\nu} - \phi^- l^+)$,
$(u_L d_L) d_R (\bar{\phi}^0 \bar{\nu} - \phi^- l^+)$
are dimension-seven~\cite{l14,bw15,lm16}.
Note that $n \to \pi^+ e^-$ is possible, but it requires the conversion of
$\phi^+$ to $\pi^+$, so it is very much suppressed compared to $n \to \pi^0
\nu$ and $p \to \pi^+ \nu$.

\noindent \underline{\it Evaluations of the loop integrals}~:~\\
The evaluation of the integral involved in the one-loop diagram of Fig.~1
is well-known. The $(\lambda_5/2)(\Phi^\dagger \eta)^2 + H.c.$ interaction
splits the complex scalar $\eta^0 = (\eta_R + i \eta_I)/\sqrt{2}$ into two
mass eigenstates with different eigenvalues $m_{R,I}$, i.e.
\begin{equation}
m_R^2 - m_I^2 = 2 \lambda_5 v^2,
\end{equation}
with $v= \langle \phi^0 \rangle = 174~\textrm{GeV}$. For a given $N$
with mass $m_N$, their contribution is given by
\begin{equation}
I_1(m_N,m_R,m_I) = {m_N \over 16 \pi^2} \left[ {m_R^2 \ln (m_N^2/m_R^2)
\over m_N^2 - m_R^2} - {m_I^2 \ln (m_N^2/m_I^2) \over m_N^2 - m_I^2} \right].
\end{equation}
The analog integral for Fig.~2 is
\begin{equation}
I_2(m_N,m_\xi) = \frac{m_N}{16 \pi^2} \left[\frac{1}{m_\xi^2-m_N^2} -
\frac{m_N^2\ln(m_\xi^2/m_N^2)}{(m_\xi^2 - m_N^2)^2}\right].
\end{equation}
For Fig.~3, we assume
\begin{equation}
m^2_{12}, m^2_{R,I} < < m^2_N < < m^2_\xi \simeq m^2_{1,2},
\end{equation}
then the integral is proportional to $\mu_1 \mu_2 m_N m_{12}^2$, where
$\mu_{1,2}$ are the trilinear couplings of $\chi_1 \delta^* \xi$ and
$\chi_2 \bar{\phi}^0 \eta^0$, and the contribution
of $m^2_{R,I}$ is negligible.  We obtain
\begin{eqnarray}
I_3(m_N,m_\xi) &=& {\mu_1 \mu_2 m_N m_{12}^2 \over 16 \pi^2} \left[
{-2 m_\xi^2 + m_N^2 \over 2 m^4_\xi (m^2_\xi - m^2_N)^2} +
{\ln (m^2_\xi/m^2_N) \over (m^2_\xi - m^2_N)^3} \right] \nonumber \\
&\simeq& {\mu_1 \mu_2 m_N m_{12}^2 [-1 + \ln(m_\xi^2/m_N^2)] \over
16 \pi^2 m_\xi^6}.
\end{eqnarray}
Using the same assumption of Eq.~(5), we obtain also
\begin{equation}
I_1 \simeq {1 \over 16 \pi^2} \left[ {m_R^2 \over m_N} \ln \left(
{m_N^2 \over m_R^2} \right) - {m_I^2 \over m_N} \ln \left(
{m_N^2 \over m_I^2} \right) \right], ~~~ I_2 \simeq {m_N \over 16 \pi^2 m^2_\xi}.
\end{equation}
The integral $I_4$ for Fig.~4 is helicity suppressed, so that $m_N$ has to be
replaced by $m_d$, hence it is negligible compared to $I_3$ and will not be
considered further.  There are also contributions to $I_{1,2}$ from
$\chi_{2,1}$, but they are suppressed by $\mu_{2,1}^2/m_{2,1}^2$.

\noindent \underline{\it Phenomenological details}~:~\\
The $3 \times 3$ neutrino mass matrix is given by
\begin{equation}
({\cal M}_\nu)_{ij} = \sum_{k} h_{ik} h_{jk} I_1(m_{N_k},m_R,m_I),
\end{equation}
where $h$ are the Yukawa $\bar{N}_R \nu \eta^0$ couplings.
The applicability of this formula has been studied extensively.
For example, if $h \sim 10^{-3}$, $m_{R,I} \sim 100$ GeV, and
$m_N \sim 10^6$ GeV, then neutrino masses are of order 0.1 eV.

The above mechanism has also the built-in possibility~\cite{m06-1} of
leptogenesis~\cite{fy1986} from the decay of $N \to l^\pm \eta^\mp$.
In particular, the required CP asymmetry can obtain a resonantly enhanced
contribution from self-energy corrections~\cite{fps1995} since the decaying
singlet fermions may have a quasi-degenerate mass spectrum, i.e.
\begin{equation}
\varepsilon_{N_i}^{}=\frac{1}{8\pi}\sum_{j\neq i}\frac{\textrm{Im}
\{[(h^\dagger h)_{ij}]^2_{}\}}{(h^\dagger h)_{ii}}
\frac{m_{N_j}m_{N_i}}{m_{N_j}^2-m_{N_i}^2}\,.
\end{equation}
Consider as usual the quantity
\begin{equation}
K_i=\frac{\Gamma_{N_i}}{2H(T)}\left|_{T=m_{N_i}}\right. ~~\textrm{with}~~
\Gamma_{N_i}^{}=\frac{1}{8\pi}(h^\dagger h)_{ii}m_{N_i}\,,~~H(T)=\left
(\frac{8\pi^{3}g_{\ast}}{90}\right)^{\frac{1}{2}}\frac{T^{2}}{M_{Pl}}\,.
\end{equation}
As an example, let $h \sim 10^{-3}$ and $m_{N_3} \gg m_{N_{1,2}} \sim 10^6\,
\textrm{GeV}$.  We then obtain the CP asymmetries $\varepsilon_{N_{1,2}}=
\mathcal{O}(0.01-0.1)$ for $m_{N_2}-m_{N_1}=\mathcal{O}(1-10\,\textrm{GeV})$.
Using $g_\ast^{}\simeq 100$ and $M_{Pl}\simeq 10^{19}\,\textrm{GeV}$, we find
$K_{1,2}=\mathcal{O}(10^{5})$, hence $z_{1,2}\simeq 4.2(\ln K_{1,2})^{0.6}\simeq
\mathcal{O}(18)$~\cite{kt1990}. This means that the lighter singlet fermions
$N_{1,2}$ can efficiently decay to generate a lepton asymmetry at a temperature
around $T_{1,2}\simeq m_{N_{1,2}}^{}/z_{1,2} = \mathcal{O}(10^5\,\textrm{GeV})$
where the sphaleron processes are still active. The final baryon asymmetry,
which is conveniently described by the ratio of the baryon number density
$n_B$ over the entropy density $s$, is then $n_B^{}/s\simeq \varepsilon_{N_{1}}/
(g_\ast K_{1} z_{1})+\varepsilon_{N_{2}}/(g_\ast K_{2} z_{2})=\mathcal{O}
(10^{-10}_{})$ as desired~\cite{kt1990}.

As for the topic of $n-\bar{n}$ oscillation, there has been a recent
resurgence of interest~\cite{p16}.  Let the effective Hamiltonian density
be given by
\begin{equation}
{\cal H}_{eff} = \sum_i c_i {\cal O}_i
\end{equation}
where ${\cal O}_i$ are the dimension-nine operators responsible for this
transition.  Then
\begin{equation}
\langle \bar{n} | {\cal H}_{eff} | n \rangle = \sum_i c_i \langle
\bar{n} | {\cal O}_i | n \rangle \simeq \sum_i c_i \Lambda_{QCD}^6
\simeq \sum_i c_i (180~{\rm MeV})^6.
\end{equation}
Let the $d_R N_R \xi^*$ coupling be $f_\xi$, the $u_{L,R} d_{L,R} \delta$
couplings be $f_\delta^{L,R}$, and the $(\delta^* \xi)^2$ coupling be
$\lambda/2$, then
\begin{equation}
\sum_i c_i = {(f_\delta^L + f_\delta^R)^2 \lambda f_\xi^2 m_N \over
16 \pi^2 m_\xi^2 m_\delta^4}.
\end{equation}
For $\tau_{n-\bar{n}} = 2 \times 10^8$ s, this translates to~\cite{p16}
\begin{equation}
\sum_i c_i = 10^{-28}~{\rm GeV}^{-5}.
\end{equation}
Inside a nucleus, the $n-\bar{n}$ transition is exponentially suppressed.
Hence the present experimental limit~\cite{olive2014}
$\tau_{n-\bar{n}} > 0.86 \times 10^8$ s yields a deuteron stability
lifetime $> 10^{31}$ y.  To match Eq.~(13) with Eq.~(14),
we may for example take again $m_N \sim 10^6$ GeV, then choose
$m_\xi \sim 10^7$ GeV, $m_\delta \sim 10^4$ GeV, and
$f_\delta^{L,R} \sim \sqrt{\lambda} \sim f_\xi \sim 0.4$.

For proton decay, the dominant decay $p \to \pi^+ \nu$ has an effective
coupling given by
\begin{eqnarray}
G_{eff} &=& {\sqrt{(f_\delta^{L})^2 + (f_\delta^R)^2} f_\xi h v I_3
\Lambda_{QCD}^3 \over f_\pi m_\delta^2} \nonumber \\
&=& \left( {\sqrt{(f_\delta^{L})^2+(f_\delta^R)^2} f_\xi h v
\Lambda_{QCD}^3 \over f_\pi m_\delta^2} \right)
{\mu_1 \mu_2 m_N m_{12}^2 [-1+\ln(m_\xi^2/m_N^2)] \over 16 \pi^2 m_\xi^6}.
\end{eqnarray}
Let $m_\xi \sim 10^7$ GeV, $m_N \sim 10^6$ GeV, $m_\delta \sim 10^4$ GeV,
$\Lambda_{QCD} = 180$ MeV, $f^{L,R}_\delta \sim f_\xi \sim 0.4$, and
$h \sim 10^{-3}$ as before.  Using $f_\pi = 130$ MeV, and choosing
$\mu_{1,2} \sim 10^5$ GeV, $m^2_{12} \sim 10^7~{\rm GeV}^2$
in addition, then $G_{eff} \sim 4.0 \times 10^{-32}$ which yields
a proton decay lifetime $\sim 1.4 \times 10^{33}$ y, using
\begin{equation}
\Gamma_p = {G_{eff}^2 (m_p^2 - m_\pi^2)^2 \over 32 \pi m_p^3}.
\end{equation}

The numerical values of the various parameters are of course for illustration
only. They are chosen to demonstrate that realistic solutions exist for
neutrino mass, neutron-antineutron oscillation, and proton decay, all in
the scotogenic context.  Again our framework assumes the validity of matter
parity $P_M$ which translates to dark parity $P_D = P_M (-1)^{2j}$, and
is derivable from lepton parity $P_L$ and baryon parity $P_B$, both of which
are respected by all dimension-four and dimension-three terms of our
renormalizable Lagrangian.  A unique dimension-two term breaks both $P_L$
and $P_B$, but preserves the product $P_M = P_L P_B$.   In the illustrative
example shown, the heaviest particles are the scalar $\xi$ and perhaps also
the scalars $\chi_{1,2}$ at $\sim 10^7$ GeV.  They decay to the singlet
fermions $N$ with mass $\sim 10^6$ GeV, which also couple to leptons and
are responsible for generating a lepton asymmetry of the Universe.
Leaving aside these very heavy particles, our proposal also predicts
a scalar diquark $\delta$ of mass $\sim 10$ TeV, as compared with the
present experimental lower limit~\cite{cms16} of about 6 TeV.  Finally, we
also have the dark scalar doublet $(\eta^+,\eta^0)$ which should be observable
at the electroweak scale.

\noindent \underline{\it Concluding remarks}~:~\\
In this scotogenic worldview, new physics phenomena beyond the standard model
are all interconnected through dark matter and dictated by the extension of
the discrete symmetries lepton parity $P_L$ and baryon parity $P_B$, both
of which are respected by the dimension-four and dimension-three terms of
our complete renormalizable Lagrangian.  A unique dimension-two term
breaks $P_L$ and $P_B$, but preserves matter parity $P_M = P_L P_B$.
Dark parity is then simply $P_M (-1)^{2j}$.

The new particles of this scenario are three dark singlet neutral Majorana
fermions $N$, a dark scalar doublet $(\eta^+,\eta^0)$, a scalar diquark
$\delta$, a dark scalar leptoquark $\xi$, and two dark real scalar singlets
$\chi_{1,2}$.  Scotogenic radiative neutrino masses are obtained through
$N$ and $\eta^0$ as shown in Fig.~1.  Leptogenesis is facilitated by
the decay $N \to l^\pm \eta^\mp$.  Neutron-antineutron oscillation is
obtained through $N$, $\xi$, and $\delta$ as shown in Fig.~2.  Slicing the
two diagrams and joining them together with $\chi_{1,2}$, proton decay is
obtained as shown in Fig.~3.  This new notion of physics beyond the standard
model connects all four fundamental processes through dark matter.
Possible experimental verfication includes the discovery of the dark
scalar doublet $(\eta^+,\eta^0)$, the crucial heavy scalar diquark $\delta$
if kinematically possible, and the prediction that proton decay should
be $p \to \pi^+ \nu$, and not $p \to \pi^0 e^+$ or $p \to \pi^+ \bar{\nu}$.

\noindent \underline{\it Acknowledgement}~:~
P.H.G. was supported in part by the Shanghai Jiao Tong University under
Grant No. WF220407201, the Recruitment Program for Young Professionals
under Grant No. 15Z127060004 and the Shanghai Laboratory for Particle
Physics and Cosmology under Grant No. 11DZ2260700. E.M. was supported
in part by the U.~S.~Department of Energy under Grant No. DE-SC0008541.

\newpage

\baselineskip 18pt
\bibliographystyle{unsrt}

\begin{thebibliography}{99}
\bibitem{m15} E. Ma, Phys. Rev. Lett. {\bf 115}, 011801 (2015).
\bibitem{m06} E. Ma, Phys. Rev. {\bf D73}, 077301 (2006).
\bibitem{v95} F. Vissani, Phys. Rev. {\bf D52}, 4245 (1995).
\bibitem{bm12} K. S. Babu and R. N. Mohapatra, Phys. Rev. Lett. {\bf 109},
091803 (2012).
\bibitem{fy1986} M. Fukugita and T. Yanagida, Phys. Lett. {\bf B174}, 45 (1986).
\bibitem{krs1985}
V.A. Kuzmin, V.A. Rubakov, and M.E. Shaposhnikov, Phys. Lett. {\bf B155},
36 (1985).
\bibitem{dm78} N. G. Deshpande and E. Ma, Phys. Rev. {\bf D18}, 2574 (1978).
\bibitem{atyy14} A. Arhrib, Y.-L. S. Tsai, Q. Yuan, and T.-C. Yuan, JCAP
{\bf 1406}, 030 (2014).
\bibitem{ikr15} A. Ilnicka, M. Krawczyk, and T. Robens, Phys. Rev. {\bf D93},
055026 (2016).
\bibitem{dku16} M. A. Diaz, B. Koch, and S. Urrutia-Quiroga, AHEP {\bf 2016},
8278375 (2016).
\bibitem{l14} L. Lehman, Phys. Rev. {\bf D90}, 125023 (2014).
\bibitem{bw15} S. Bhattacharya and J. Wudka, arXiv:1505.05264 [hep-ph].
\bibitem{lm16} Y. Liao and X.-D. Ma, arXiv:1607.07309 [hep-ph].
\bibitem{m06-1} E. Ma, Mod. Phys. Lett. {\bf A21}, 1777 (2006).

\bibitem{fps1995}
M. Flanz, E.A. Paschos, and U. Sarkar, Phys. Lett. B \textbf{345},
248 (1995).


\bibitem{kt1990}
E.W. Kolb and M.S. Turner, \textit{The Early Universe},
Addison-Wesley, 1990.


\bibitem{p16} D. G. Phillips II {\it et al.}, Phys. Rep. {\bf 612}, 1 (2016).
\bibitem{olive2014} K.A. Olive {\it et al.}, (Particle Data Group
Collaboration), Chin. Phys. {\bf C38}, 090001 (2014).
\bibitem{cms16} V. Khachatryan {\it et al.} (CMS Collaboration), Phys. Rev.
Lett. {\bf 116}, 071801 (2016).

\end{thebibliography}

\end{document}